\documentclass[11pt]{article}
\usepackage{amssymb}
\usepackage{amsmath}
\usepackage{graphicx}
\begin{document}

\begin{titlepage}
\begin{center}

\vspace*{10mm}

{\LARGE\bf
Spontaneous Gauge Symmetry Breaking in a Non-Supersymmetric D-brane Model}

\vspace*{20mm}

{\large
Noriaki Kitazawa\footnote[1]{kitazawa@phys.se.tmu.ac.jp}
 and Shutaro Kobayashi\footnote[2]{kobayashi-shutaro@ed.tmu.ac.jp}
}
\vspace{6mm}

{\it
Department of Physics, Tokyo Metropolitan University,\\
Hachioji, Tokyo 192-0397, Japan\\
}

\vspace*{15mm}

\begin{abstract}
We investigate a D-brane system
 in which spontaneous gauge symmetry breaking occurs.
The system consists of four D3-branes and three anti-D7-branes
 at $\mathbb{C}^{3}/\mathbb{Z}_{3}$ singularity.
The singularity is blown up
 by vacuum expectation values of the states of twisted closed string,
 whose effects appear as Fayet-Iliopoulos terms
 in low-energy effective field theory.
We derive an effective potential
 for scalar fields in four-dimensional effective field theory
 using the technique of superstring world-sheet theory.
Since there is no supersymmetry in this system,
 one-loop masses for some scalar fields are considered.
We find some stationary and stable vacua in which the original
 U$(2) \times $U$(1)_1 \times $U$(1)_2 \times $U$(3)$ gauge symmetry
 is spontaneously broken to
 U$(1) \times $U$(1)' \times $U$(3)$ gauge symmetry.
Both Fayet-Iliopoulos terms and one-loop masses 
 are necessary for the existence of such vacua.
We carefully investigate
 a geometrical aspect of the gauge symmetry breaking.
We show that
 the gauge symmetry breaking is understood as the separation of D3-branes.
Some implications to the possibility of low-scale string models
 are also discussed.
\end{abstract}

\end{center}
\end{titlepage}

\section{Introduction}

It is believed that superstring theory is a possible framework
 which naturally includes quantum gravity.
There are five possible types of superstring theories in flat space-time.
In view of unified theory,
 one expects that all these five theories may be described by
 a single M-theory with eleven-dimensional space-time.

Superstring theory is also expected
 to describe the physics beyond the standard model.
Heterotic string theory or type II theory with multiple D-branes
 can be used to construct phenomenological models.
In particular,
 in order to construct realistic models in type II string theory,
 we need some special D-brane configurations:
 D-branes at singularities, intersecting D-branes, magnetized D-branes,
 and so on
 (see ref.\cite{Blumenhagen:2006ci}, for example).
In general, gauge couplings are described by the string coupling $g_{s}$,
 the Regge slope parameter $\alpha'$,
 and partial volumes of compactified spaces.

As a possibility of realistic superstring models,
 there is a scenario of low-scale string models (or TeV-scale string models) 
 with large-volume compactifications
 \cite{Antoniadis:1990ew,Antoniadis:1998ig,Cicoli:2011yy}.
Low-scale string models can not be realized in heterotic string theory,
 since gauge couplings become too small under large-volume compactifications.
On the other hand,
 type II string theory can describe low-scale string models using D-branes,
 since gauge couplings can have reasonable values
 with appropriate configurations of D-branes on which
 gauge symmetries are realized as open-string degrees of freedom.
Note that the low-scale string itself
 is not a solution of the hierarchy problem.
There must be a natural mechanism for electroweak symmetry breaking
 with correct energy scale.

For low-scale string models,
 it is important to investigate D-brane models without supersymmetry,
 because natural electroweak symmetry breaking can take place
 through non-trivial radiative corrections to Higgs potential.
In low-scale string model,
 it is natural to break supersymmetry by D-brane configurations.
In the system of D-branes at singularities,
 one way to obtain non-supersymmetric models
 is to introduce anti-D-branes at some singularities
 \cite{Aldazabal:2000sa}.
We obtain different massless spectra
 between bosons and fermions in such systems.
There must be non-trivial mass corrections at one-loop,
 since there is no supersymmetry \cite{Antoniadis:2000tq,Kitazawa:2008tb}.

In this paper we investigate the gauge symmetry breaking
 and its geometrical aspects in a model without supersymmetry.
We would like to emphasize that
 the aim of this work is not to construct a concrete model
 with spontaneous gauge symmetry breaking by some D-brane dynamics,
 but investigate the phenomena with possible assumptions
 paying special attention to geometrical understanding.
The dynamics of spontaneous gauge symmetry braking in string theory
 is rarely investigated, though it must be interesting
 in both phenomenological and theoretical
 (as a problem of D-brane dynamics) point of view.

This paper is organized as follows.
In the next section,
 we introduce a system of type IIB string theory
 with four D3-branes and three anti-D7-branes
 at $\mathbb{C}^{3}/\mathbb{Z}_{3}$ singularity.
The model satisfies (twisted) Ramond-Ramond tadpole cancellation conditions,
 and it is a quite simple model with non-Abelian gauge symmetry.
We obtain the potential of D3-brane moduli fields
 through the calculations of four-point amplitudes
 in world-sheet conformal field theory.
In section 3,
 we discuss the vacuum structure,
 which is described by the vacuum expectation values of D3-brane moduli fields,
 with geometrical understanding.
D3-brane moduli fields have a flat direction
 on which gauge symmetry is spontaneously broken in general,
 and this vacuum structure can be geometrically understood
 as the separation of four D3-branes.
We also investigate
 the deformation of this vacuum structure (flat direction)
 by the effect of Fayet-Iliopoulos terms
 which are controlled by the vacuum expectation values
 of twisted moduli fields from twisted closed string.
These vacuum expectation values result
 various ways of blowing-up the orbifold singularity
 $\mathbb{C}^{3}/\mathbb{Z}_{3}$.
In section 4,
 we further include one-loop mass corrections to D3-brane moduli fields.
The vacuum structure is deformed again,
 and all the flat directions are lifted up.
This can be geometrically understood that
 the position of D3-branes are stabilized and fixed.
There are some stationary and stable points
 in the effective potential of D3-brane moduli fields,
 at which spontaneous gauge symmetry breaking occurs.
We discuss the relation
 between the scale of gauge symmetry breaking and string scale,
 which gives some implications to the possibility of low-scale string models.
In section 5, we summarize our results.

\section{D3-branes and anti-D7-branes
 at $\mathbb{C}^{3}/\mathbb{Z}_{3}$ singularity}

Consider a stack of four D3-branes and a stack of three anti-D7-branes
 in flat ten-dimensional space-time of type IIB theory.
Take the coordinates of the world-volume of the D3-brane stack
 $X^\mu$ with $\mu=0,1,2,3$,
 and take those of the anti-D7-brane stack $X^M$ with $M=0,1,\cdots,7$.
This system gives U$(4) \times $U$(3)$ gauge symmetry
 with massless matter fields in adjoint representations,
 if these two D-brane stacks are separated in $(X^{8},X^{9})$-plane.

Next, consider a $\mathbb{Z}_{3}$ symmetry with twist vector
 $(1,1,-2)$ in the six-dimensional space outside D3-brane world-volume,
 and apply a projection under this symmetry to obtain
 matter fields in fundamental representations of the gauge group
 \cite{Aldazabal:2000sa}.
We introduce an index $i=1,2,3$
 which labels three complex coordinates ($\mathbb{C}^{3}$)
 of the six-dimensional space
\begin{equation}
 Z^{i} = \frac{1}{\sqrt{2}}\left(X^{2i+2}+iX^{2i+3}\right).
\end{equation}
We assume that
 the D3-branes stack is set at the origin of this six-dimensional space,
 and the anti-D7-brane stack is set at the origin of $(X^{8},X^{9})$-plane.
The $\mathbb{Z}_{3}$ twist operator
 to the Chan-Paton factor of the open string on D3-branes is set as
\begin{equation}
\gamma_{3}=\textnormal{diag}(1_{2\times 2},\alpha,\alpha^{2}),
\end{equation}
 and the $\mathbb{Z}_{3}$ twist operator
 to that of the open string on anti-D7-branes is set as
\begin{equation}
\gamma_{7}=1_{3\times 3},
\end{equation}
 where $\alpha \equiv \textnormal{exp}\left(2\pi i/3\right)$.
After the projection,
 the gauge symmetry on D3-brane stack is broken
 from U$(4)$ to U$(2) \times$U$(1)_1 \times$U$(1)_2$,
 and the total remaining gauge symmetry becomes
\begin{equation}
 {\rm U}(2) \times {\rm U}(1)_1 \times {\rm U}(1)_2 \times {\rm U}(3).
\end{equation}
The massless four-dimensional scalar fields of ``33 sector'',
 which come from the open string whose both edges are on the D3-brane stack,
 are
\begin{equation}
 A_{i}^{(1)} : (2,0,-1,1)\ ,\
 A_{i}^{(2)} : (\bar{2},+1,0,1)\ ,\ 
 A_{i}^{(3)} : (1,-1,+1,1),
\label{33-sector}
\end{equation}
 where the index $i=1,2,3$ has been introduced above,
 and those of ``37 sector'',
 which come from the open string whose one edges is on the D3-brane stack
 and another edge is on the anti-D7-brane stack,
 are  
\begin{equation}
 \phi^{(3\bar{7})}_{1} : (2,0,0,\bar{3})\ ,\ 
 \phi^{(3\bar{7})}_{2} : (\bar{2},0,0,3).
\label{37-sector}
\end{equation}
The massless four-dimensional fermion fields in 33 sector
 form nine chiral supermultiplets with nine complex scalar fields
 in eq.(\ref{33-sector}).
On the other hand,
 the massless four-dimensional fermion fields in 37 sector
 belong to different representations from two complex scalar fields
 in eq.(\ref{37-sector}).
Four-dimensional supersymmetry is broken in the spectrum by construction.
In this system
 all the twisted Ramond-Ramond tadpoles are canceled,
 and there is no gauge anomaly.
Uncanceled untwisted Ramond-Ramond tadpole is formally harmless
 in the system without compactification.
Although uncanceled NS-NS tadpoles, where NS stands for Neveu-Schwarz,
 indicates that the present flat background metric
 is not the solution of string theory\cite{Fischler:1986ci,Fischler:1986tb},
 we assume that the possible modification to the metric
 would be small enough so that it could be treated perturbatively.

The potential for 33 sector scalar fields
 can be obtained through the $\mathbb{Z}_{3}$ projection of the potential
 in U$(2)\times $U$(1)_1\times $U$(1)_2$ super-Yang-Mills theory
 using the four-dimensional $\mathcal{N}=1$ superspace formalism,
 since D3-brane world-volume theory without 37 strings
 has four-dimensional supersymmetry.
There are two contributions of F-term and D-term potentials:
\begin{equation}
 V_{33} = V_{F} + V_{D}.
\label{DD_potential}
\end{equation}
The F-term potential $V_{F}$ is given by
\begin{equation}
 V_{F} = |F_{i}^{(1)}|^{2} + |F_{i}^{(2)}|^{2} + |F_{i}^{(3)}|^{2}
\end{equation}
 with
\begin{equation}
 F_{i}^{(1)\dag} = g \epsilon_{ijk} A_{j}^{(3)} (A_{k}^{(2)})^T\ ,\ 
 F_{i}^{(2)\dag} = g \epsilon_{ijk} (A_{j}^{(1)})^T A_{k}^{(3)}\ ,\ 
 F_{i}^{(3)\dag} = g \epsilon_{ijk} (A_{j}^{(2)})^T A_{k}^{(1)},
\end{equation}
 where $g$ is the gauge coupling constant.
The D-term potential $V_{D}$ is given by 
\begin{equation}
 V_{D} = \frac{1}{2}(D_{U(2)}^{\hat{a}})^{2}
       + \frac{1}{2}D_{1}^{2}
       + \frac{1}{2}D_{2}^{2}
\end{equation}
 with
\begin{eqnarray}
 D_{U(2)}^{\hat{a}}
  &=& - g \left( A_{i}^{(1)\dag}T^{\hat{a}}A_{i}^{(1)}
               - A_{i}^{(2)\dag}(T^{\hat{a}})^*A_{i}^{(2)} \right),\\
 D_{1}
  &=& - \frac{g}{\sqrt{2}}
          \left( |A_{i}^{(2)}|^{2} - |A_{i}^{(3)}|^{2} \right),\\
 D_{2}
  &=& - \frac{g}{\sqrt{2}}
          \left( |A_{i}^{(3)}|^{2} - |A_{i}^{(1)}|^{2} \right),
\end{eqnarray}
 where $\hat{a} = 0,1,2,3$ and $T^{\hat{a}}$ are U$(2)$ generators
\begin{equation}
 T^{0} = \frac{1}{2}\left(\begin{array}{rr}1&0\\ 0&1\end{array}\right),
\quad
 T^{1} = \frac{1}{2}\left(\begin{array}{rr}0&1\\ 1&0\end{array}\right),
\quad
 T^{2} = \frac{1}{2}\left(\begin{array}{rr}0&-i\\ i&0\end{array}\right),
\quad
 T^{3} = \frac{1}{2}\left(\begin{array}{rr}1&0\\ 0&-1\end{array}\right).
\end{equation}
We describe generators of two U$(1)$ symmetries,
 U$(1)_1$ and U$(1)_2$, as $Q_1$ and $Q_2$, respectively.
In the above equations
 the summations for repeated indices $i,j,k$
 and hidden gauge indices are understood, for example:
\begin{equation}
 |A^{(3)}_i|^2
  = \sum_{i=1}^3 \left(A^{(3)}_i\right)^\dag A^{(3)}_i,
\qquad
 |A^{(1)}_i|^2
  = \sum_{\alpha=1}^2 \sum_{i=1}^3
    \left(A^{(1)}_i\right)^\dag_\alpha
    \left(A^{(1)}_i\right)^\alpha,
\end{equation}
 where $\alpha$ is the index of fundamental representation of SU$(2)$.

We need to calculate some disc amplitudes
 in world-sheet conformal field theory
 to obtain the potential which includes both 33 and 37 sector scalar fields.
The explicit form of the vertex operators for 33 and 37 scalar states
 should be required with careful determination of their normalization
 through the calculations of disc amplitudes with gauge bosons.
We should introduce two types of vertex operators
 called picture $-1$ and picture $0$.
The picture $-1$ vertex operators for 37 scalar states
 are obtained as \cite{Hashimoto:1996he}
\begin{equation}
 \phi_1^{(3\bar{7})} \,
  : \lambda^{(3\bar{7})} S e^{-\phi}\Delta e^{ik\cdot X},
\quad
 \phi_2^{(3\bar{7})} \,
  : \lambda^{(3\bar{7})\dag} S e^{-\phi}\Delta e^{ik\cdot X},
\label{picture_-1}
\end{equation}
 where $\lambda^{(3\bar{7})}$ describes Chan-Paton degrees of freedom
 which labels the D-branes on which the edges of string end,
 and the spin field is defined by bosonization
\begin{equation}
 S = \exp\left(\frac{i}{2}H_{1}-\frac{i}{2}H_{2}\right)
\end{equation}
 with the operator product expansion (OPE) of between boson fields
\begin{equation}
 H_{i}(x_{1})H_{j}(x_{2})
  = \delta_{ij}\textnormal{ln}(x_{1}-x_{2})+\textnormal{regular}.
\end{equation}
The operator $\Delta$
 represents Dirichlet-Neumann boundary condition
 in the space of $X^M$ with $M=4,5,6,7$,
 and it is described by the product of
 four boundary condition changing operators for each direction
\begin{equation}
 \Delta = \sigma^{4}\sigma^{5}\sigma^{6}\sigma^{7}.
\end{equation}
The algebraic structure of the boundary condition changing operators,
 in particular, their OPE rule,
 is given in refs.\cite{Hashimoto:1996he,Zamolodchikov:1987ae}.
The picture $0$ vertex operators for 33 scalar states are obtained as
\begin{equation}
 A_{j}
  = \lambda \frac{1}{\sqrt{2\alpha'}}
    \left(
     i\partial X^{(-)}_j + 2\alpha'k\cdot \psi\psi^{(-)}_j 
    \right) e^{ik\cdot X}\ ,\ 
\label{picture_0}
\end{equation}
 where $j=1,2,3$ with
\begin{equation}
 X^{(\pm)}_j
  = \frac{1}{\sqrt{2}} \left( X^{2j+2} \pm iX^{2j+3} \right),
\quad
 \psi^{(\pm)}_j
  = \frac{1}{\sqrt{2}} \left( \psi^{2j+2} \pm i\psi^{2j+3} \right),
\end{equation}
 and $\lambda$ describes Chan-Paton degrees of freedom.
We should choose appropriate $\lambda$
 to describe three kinds of scalar fields,
 $A^{(1)}$, $A^{(2)}$ and $A^{(3)}$,
 with different gauge quantum numbers.
Note that the indices of inner products with four-momentum $k_\mu$
 in eqs.(\ref{picture_-1}) and (\ref{picture_0})
 are running over $\mu=0,1,2,3$.

Using these vertex operators and calculating disc amplitudes,
 the potential with 33 and 37 scalar fields are obtained as
\begin{equation}
\begin{split}
V_{\rm mixed}
 = - \frac{g^{2}}{4}
     A_{1}^{(1)\dag}
      \left(
       \phi^{(3\bar{7})}_{1}\phi^{(3\bar{7})\dag}_{1}
       + \phi^{(3\bar{7})\dag}_{2}\phi^{(3\bar{7})}_{2}
      \right)
     A_{1}^{(1)}
   + \frac{g^{2}}{4}
     A_{1}^{(2)\dag}
      \left(
       \phi^{(3\bar{7})}_{1}\phi^{(3\bar{7})\dag}_{1}
       + \phi^{(3\bar{7})\dag}_{2}\phi^{(3\bar{7})}_{2}
      \right)
     A_{1}^{(2)}\\
   + \frac{g^{2}}{4}
     A_{2}^{(1)\dag}
      \left(
       \phi^{(3\bar{7})}_{1}\phi^{(3\bar{7})\dag}_{1}
       + \phi^{(3\bar{7})\dag}_{2}\phi^{(3\bar{7})}_{2}
      \right)
     A_{2}^{(1)}
   - \frac{g^{2}}{4}
     A_{2}^{(2)\dag}
      \left(
       \phi^{(3\bar{7})}_{1}\phi^{(3\bar{7})\dag}_{1}
       + \phi^{(3\bar{7})\dag}_{2}\phi^{(3\bar{7})}_{2}
      \right)
     A_{2}^{(2)}\\
   + \frac{g^{2}}{4}
     A_{3}^{(1)\dag}
      \left(
       \phi^{(3\bar{7})}_{1}\phi^{(3\bar{7})\dag}_{1}
       + \phi^{(3\bar{7})\dag}_{2}\phi^{(3\bar{7})}_{2}
      \right)
     A_{3}^{(1)}
   + \frac{g^{2}}{4}
     A_{3}^{(2)\dag}
      \left(
       \phi^{(3\bar{7})}_{1}\phi^{(3\bar{7})\dag}_{1}
       + \phi^{(3\bar{7})\dag}_{2}\phi^{(3\bar{7})}_{2}
      \right)
     A_{3}^{(2)}.
\end{split}
\label{mixed_potential}
\end{equation}
It is interesting to see that
 the masses of some 37 scalar fields
 by the equal vacuum expectation values of
 $A^{(1)}_1$ and $A^{(2)}_1$ or $A^{(1)}_2$ and $A^{(2)}_2$ are canceled,
 but, on the other hand,
 the masses of some 37 scalar fields
 by the equal vacuum expectation values of
 $A^{(1)}_3$ and $A^{(2)}_3$ are not canceled.

The potential for 37 sector scalar fields is constrained
 by considering trace structure of disc diagrams
 with four vertex operators on the boundary,
 gauge invariance, and H-momentum conservation
 \footnote{
 In world-sheet theories as two-dimensional conformal field theories,
 fermionic fields can be represented by bosonic fields such that
 $\psi(x)=e^{iaH(x)}$ with some constants, $a\neq 0$.
 The H-momentum conservation is a condition that
 the $n$-point correlation functions $\langle\psi_{1}\cdots\psi_{n}\rangle$
 vanish if $a_{1}+\cdots +a_{n} \ne 0$.}.
The form of the potential should be
\begin{equation}
\begin{split}
 V_{37} =
  \kappa_{1}g^{2}
   \textnormal{Tr}
   \left(
    \phi^{(3\bar{7})\dag}_{1}\phi^{(3\bar{7})}_{1}
    \phi^{(3\bar{7})\dag}_{1}\phi^{(3\bar{7})}_{1}
   \right)
 +\kappa_{2}g^{2}
   \textnormal{Tr}
   \left(
    \phi^{(3\bar{7})}_{2}\phi^{(3\bar{7})\dag}_{2}
    \phi^{(3\bar{7})}_{2}\phi^{(3\bar{7})\dag}_{2}
   \right)\\
 +\kappa_{3}g^{2}
   \textnormal{Tr}
   \left(
    \phi^{(3\bar{7})\dag}_{1}\phi^{(3\bar{7})}_{1}
    \phi^{(3\bar{7})}_{2}\phi^{(3\bar{7})\dag}_{2}
   \right)
 +\kappa_{4}g^{2}
   \textnormal{Tr}
   \left(
    \phi^{(3\bar{7})}_{1}\phi^{(3\bar{7})\dag}_{1}
    \phi^{(3\bar{7})\dag}_{2}\phi^{(3\bar{7})}_{2}
   \right),
\end{split}
\end{equation}
 where $\kappa$'s are some constants
 whose values are not estimated in this paper.
The values of $\kappa$'s depend on the ways of compactifications
 and this is an effort to construct a specific model
 which is not the intention of this paper.
We investigate what happens on D3-brane world-volume
 by simply assuming that all the 37 sector fields are stabilized
 with zero vacuum expectation value,
 $\langle \phi_{1}^{(3\bar{7})} \rangle
  = \langle \phi_{2}^{(3\bar{7})} \rangle= 0$.

\section{Geometrical understanding of the vacuum structure}

In this section,
 we investigate the vacuum structure of the potential of eq.(\ref{DD_potential}) 
 and its deformation by Fayet-Iliopoulos (FI) terms
 which are controlled by vacuum expectation values of twisted moduli fields.

The potential of eq.(\ref{DD_potential}) has flat directions:
\begin{equation}\label{flat direction}
 A^{(1)}_{i} = \left(\begin{array}{rr}a_{i}\\ 0\ \end{array}\right)\ ,\ 
 A^{(2)}_{i} = \left(\begin{array}{rr}b_{i}\\ 0\ \end{array}\right)\ ,\ 
 A^{(3)}_{i} = c_{i}
\end{equation}
 for arbitrary values of $a_{i}=b_{i}=c_{i}$.
The gauge symmetry
 U$(2)\times $U$(1)_1\times $U$(1)_2$
 is spontaneously broken to U$(1) \times $U$(1)'$
 at the general points on the flat directions
 except for the origin of the moduli space $a_{i}=b_{i}=c_{i}=0$.
The generators correspond to two U$(1)$ gauge symmetries,
 U$(1)$ and U$(1)'$, are
 $Q_1+Q_2+(T^0+T^3)$ and $T^0-T^3$, respectively.
In geometrical point of view,
 it looks like that two of four D3-branes are lost.
Keeping in mind that $\mathbb{C}^{3}$ is divided by $\mathbb{Z}_{3}$,
 this gauge symmetry breaking can be understood as separations of D3-branes:
 one D3-brane, which corresponds to the quantum number of $T^0-T^3$,
 stays at the origin
 and the other three D3-branes, which correspond to the quantum numbers of
 $Q_1$, $Q_2$ and $T^0+T^3$, leave from the origin
 in $\mathbb{Z}_{3}$ symmetric ways.
In this way we can understood that
 U$(1)$ is realized on these identified three D3-branes
 separated from the origin,
 and U$(1)'$ is realized on a D-brane at the origin.
A schematic picture is shown in fig.1.

This geometrical understanding is confirmed
 by the potential of eq.(\ref{mixed_potential}).
The separation in $(X^{8},X^{9})$-plane,
 which is described by the vacuum expectation value of $a_3$,
 stretches the open string of 37 sector,
 and some scalar fields in 37 sector should become massive.
We see from the potential of eq.(\ref{mixed_potential}) that
 the scalar fields in 37 sector
 which are charged under $T^0+T^3$ become massive,
 and those which are charged under $T^0-T^3$ remain massless.
Here, we assume that the stack of anti-D7-brane are fixed at the origin
 of $(X^{8},X^{9})$-plane
 (no vacuum expectation values for 37 sector scalar fields).
In the separation in $(X^{4},X^{5})$-plane or $(X^{6},X^{7})$-plane,
 which are described by the vacuum expectation value of $a_1$ or $a_2$,
 respectively, do not stretch the open string of 37 sector,
 and all the scalar fields in 37 sector should remain massless.
Note that both $(X^{4},X^{5})$-plane and $(X^{6},X^{7})$-plane
 are included in anti-D7-brane world-volume.
This fact is also realized in the potential of eq.(\ref{mixed_potential}).
There are many works on the related issue
 that the effective field theory on D-branes at a singularity
 describes the geometry of the singularity
 (see ref.\cite{Douglas:1997de}, for example).

\begin{figure}[t]
\begin{center}
\unitlength=0.8mm
\includegraphics[scale=0.7]{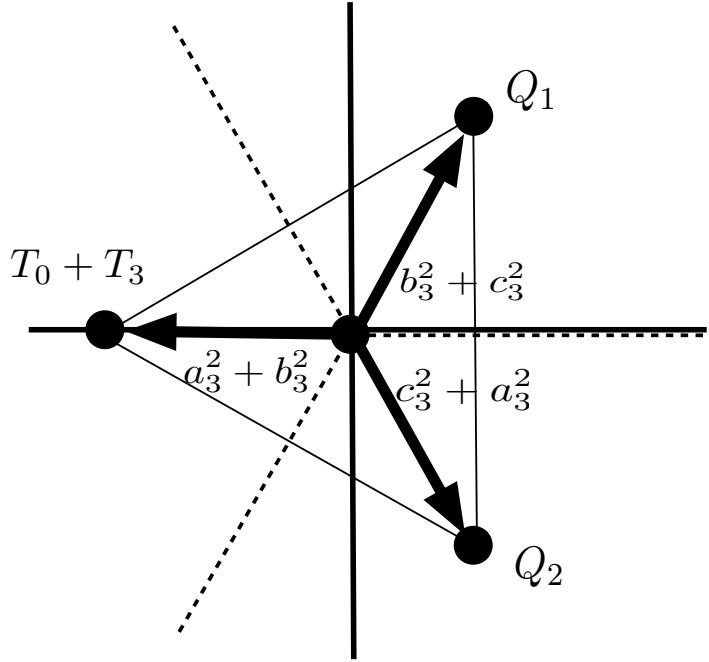} 
\caption{
A schematic picture of
 $\mathbb{Z}_{3}$ symmetric configuration of D3-branes in $(X^{8},X^{9})$-plane
 corresponding to a flat direction in the potential of eq.(\ref{DD_potential}).
The black circles indicate D3-branes.
The distances between the origin and D3-branes
 are proportional to $a_3^2+b_3^2$, $b_3^2+c_3^2$ and $c_3^2+a_3^2$
 as shown in the figure.
Three separated D3-branes form an equilateral triangle. 
}
\end{center}
\end{figure}

Next, consider the effect of FI term
 which is controlled by the vacuum expectation values of twisted moduli fields.
The possible general modification by FI terms
 in D-term potential is described as
\begin{equation}
\begin{split}
 D_{1} =
    - \frac{g}{\sqrt{2}}
    \left(
     |A_{i}^{(2)}|^{2} - |A_{i}^{(3)}|^{2} + \xi_{1}
    \right),\\
 D_{2} =
    - \frac{g}{\sqrt{2}}
    \left(
     |A_{i}^{(3)}|^{2} - |A_{i}^{(1)}|^{2} + \xi_{2}
    \right),\\
 D_{U(2)}^{\hat{a}=0} =
    - \frac{g}{2}
    \left(
     |A_{i}^{(1)}|^{2} - |A_{i}^{(2)}|^{2} + \xi_{3}
    \right),
\end{split}
\end{equation}
 where $\xi_{1}$, $\xi_{2}$ and $\xi_{3}$ are coefficients of FI terms.
In our system
 these coefficients of FI terms are determined by the value of
 a single twisted moduli field $\phi=\phi_{R}+i\phi_{I}$ as
 \cite{Douglas:1997de} 
\begin{equation}
\begin{split}
 \xi_{1} = -\frac{1}{2}\phi_{R} - \frac{\sqrt{3}}{2}\phi_{I}\ ,\ 
 \xi_{2} = -\frac{1}{2}\phi_{R} + \frac{\sqrt{3}}{2}\phi_{I}\ , \ 
 \xi_{3} = 2\phi_{R},
\end{split}
\end{equation}
 and they satisfy a constraint
\begin{equation}
 \xi_{1} + \xi_{2} + \frac{1}{2}\xi_{3} = 0.
\end{equation}
For simplicity and for the stability of 37 sector scalar fields,
 we only consider the modification of a flat direction with
 $\langle A^{(a)}_{i=1,2}\rangle = 0$
 and $A^{(a)}_{i=3} \equiv A^{(a)} \ne 0$
 with $a=1,2,3$.
In this case $V_{F}=0$ is trivially satisfied, 
 and we may consider the D-term potential $V_{D}$ only.
The mass of 37 sector scalar fields are generated by
 $A^{(1)}, A^{(2)} \ne 0$ through the potential of eq.(\ref{mixed_potential}).
It is easily to see that
 the origin of the moduli space, $A^{(a)}=0$ for all $a=1,2,3$,
 is stationary but unstable.
The vacuum energy at that point is
\begin{equation}
 V_{\rm origin} = \frac{5}{8}g^{2}\phi_{R}^{2} + \frac{3}{8}g^{2}\phi_{I}^{2}.
\label{energy_origin}
\end{equation}
We assume that twisted moduli fields, $\phi_R$ and $\phi_I$,
 are stabilized with finite vacuum expectation values less than
 string scale, $\phi_R, \phi_I < M_s^2$ with $M_s \equiv 1/\sqrt{\alpha'}$,
 so that the effect of FI terms,
 namely the effect of blowing up the orbifold singularity,
 can be treated perturbatively.
We do not specify the mechanism of the stabilization
 which is beyond the scope of this paper
 (see refs.\cite{Abel:2000tf,Higaki:2003zk}, for example).

We find a flat direction with the above assumptions.
\begin{equation}
 A^{(1)} = \left(\begin{array}{rr}a\\ 0\end{array}\right)\ ,\ 
 A^{(2)} = \left(\begin{array}{rr}b\\ 0\end{array}\right)\ ,\ 
 A^{(3)} = c
\label{ansatz}
\end{equation}
 with
\begin{equation}
 b^{2}= a^{2} + \phi_{R},
\label{solution_1}
\end{equation}
\begin{equation}
 c^{2} = a^{2} + \frac{1}{2}\phi_{R} - \frac{\sqrt{3}}{2}\phi_{I}.
\label{solution_2}
\end{equation}
Any value of $a$ is allowed as far as that
 $b^2, c^2 > 0$ are satisfied for given values of $\phi_R$ and $\phi_I$.
The vacuum energy of the flat direction is
\begin{equation}
 V_{\rm flat} = \frac{g^{2}}{4}\phi_{R}^{2},
\end{equation}
 which is always smaller than the vacuum energy of eq.(\ref{energy_origin})
 at the origin of the moduli space
 with non-zero $\phi_R$ and $\phi_I$.
It is interesting to see that
 this vacuum energy does not depend on $\phi_I$,
 and $\phi_R$ controls the magnitude of supersymmetry breaking
 on the flat direction.
We can suggest that
 $\phi_I$ gives blown-up orbifold preserving supersymmetry,
 and $\phi_R$ gives blown-up orbifold without supersymmetry.
The stability condition are also satisfied on the flat direction
 \footnote{
 The stability condition means that
 all the eigenvalues of the matrix,
 which is formed by the second order derivatives
 of the potential by fields, are not negative.}.
The gauge symmetry is spontaneously broken
 from U$(2)\times $U$(1)_1\times $U$(1)_2$ to U$(1)\times $U$(1)'$
 in general point of the flat direction with non-zero $\phi_R$.
In case of $\phi_R = 0$ and  $\phi_I < 0$,
 the flat direction includes the point of $a = b = 0$ and $c \ne 0$,
 where U$(2)$ gauge symmetry is not broken.

To obtain geometrical understanding of this flat direction
 it is easy to consider the case of $\phi_R=0$, first.
Now the flat direction is given by
\begin{equation}
 b^2 = a^2,
\qquad
 c^2 = a^2 -{\sqrt{3} \over 2} \phi_I,
\end{equation}
 where any value of $a$ is allowed as far as $c^2 > 0$
 for given value of $\phi_I$.
The gauge symmetry is spontaneously broken to
 U$(1)\times $U$(1)'$ from the original U$(2)\times $U$(1)_1\times $U$(1)_2$
 for general values of $a \ne 0$.
As a small modification of the flat directions in case without FI term,
 this flat direction should corresponds to the geometrical picture
 of that one of four D3-branes is staying at the origin
 and remaining three D3-branes are leaving from the origin
 with identification by some ``modified $\mathbb{Z}_{3}$''.
A schematic picture is shown in fig.2.
These three D3-branes do not form an equilateral triangle,
 but form an isosceles triangle.
We understand that
 this is the result of blowing-up the orbifold singularity by $\phi_I$
 to some smooth manifold.

\begin{figure}[t]
\begin{center}
\unitlength=0.8mm
\includegraphics[scale=0.5]{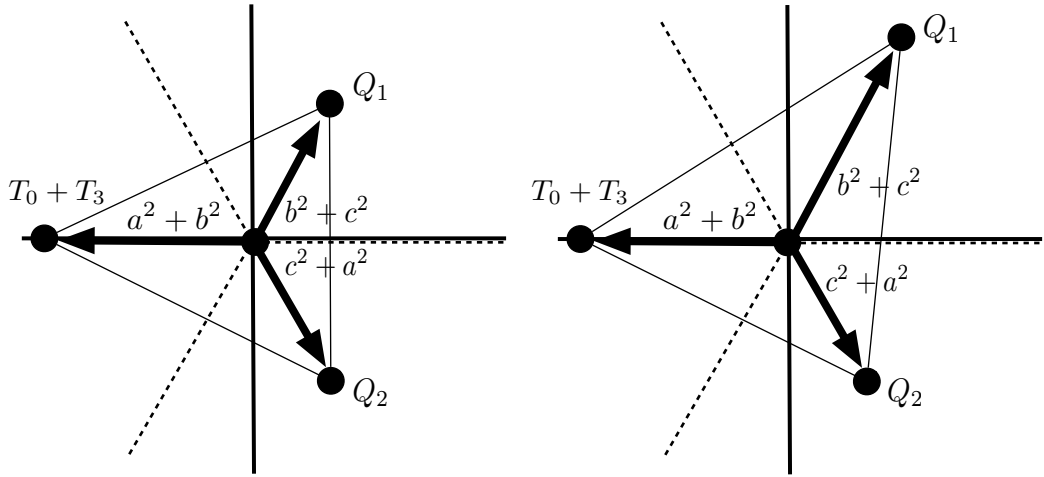} 
\caption{
Schematic pictures of
 the configurations of D3-branes in $(X^{8},X^{9})$-plane
 corresponding to the flat directions with FI terms.
Left figure corresponds to the case of $\phi_R = 0$ and $\phi_I \ne 0$,
 and right figure corresponds to the case of $\phi_R, \phi_I \ne 0$.
The black circles indicate D3-branes.
The distances between the origin and D3-branes
 are shown in figures.
}
\end{center}
\end{figure}

We can have a geometrical understanding
 in the same way for the case of $\phi_R \ne 0$.
In this case
 all the vacuum expectation values are different
 and a schematic picture is shown in fig.2.
Note that
 these schematic pictures should not be taken too much rigorously,
 because we do not pursuit the geometry of blown-up orbifold,
 which is not the aim of this paper.

\section{Spontaneous gauge symmetry breaking on D3-branes}

The one-loop mass corrections to D3-brane moduli fields should exists,
 since the system is non-supersymmetric with anti-D7-branes.
These masses can be generally described as
\begin{equation}
 V_{\rm mass} =
  m_{1}^{2}|A^{(1)}|^{2} + m_{2}^{2}|A^{(2)}|^{2} + m_{3}^{2}|A^{(3)}|^{2}.
\end{equation}
In this section we consider the potential of
\begin{equation}
 V = V_{D} + V_{\rm mass}.
\label{potential_with_mass}
\end{equation}
In the calculations of one-loop masses
 in non-supersymmetric systems using world-sheet conformal field theory,
 we encounter the infrared divergences due to the uncanceled NS-NS tadpoles.
We need to do NS-NS tadpole resummations \cite{Dudas:2004nd,Kitazawa:2008hv}
 or some other cares.
Though it is difficult to do rigorous calculations,
 it is not so difficult to have an order estimate.
In ref.\cite{Kitazawa:2008tb}
 a formula of the one-loop mass of D3-brane moduli in present system
 is explicitly given.
The formula consists of two contributions,
 one from NS-NS closed string exchange with infrared divergences
 and the other from Ramond-Ramond closed string exchange,
 which are canceled in case with supersymmetry.
We can estimate the magnitude of Ramond-Ramond contribution
 paying attention of the cancellation of its ultraviolet divergence
 with NS-NS contribution.
The result is very simple and understandable:
\begin{equation}
 m^{2} \simeq \frac{1}{\omega^{2}}\frac{g^{2}}{16\pi^{2}}M_{s}^{2},
\end{equation}
 where the factor $1/\omega^2$ is of the order of unity,
 if nothing special happens.
In the following
 we simply assume that $m_1^2 = m_2^2 = m_3^2 = m^2 > 0$.

The flat direction
 of eqs.(\ref{ansatz}), (\ref{solution_1}) and (\ref{solution_2})
 is lifted up by these mass terms.
The stationary conditions
 are not satisfied with that all of $a$, $b$ and $c$ are not zero,
 since it requires the condition $m_{1}^{2} + m_{2}^{2} + m_{3}^{2}=0$,
 which can not be satisfied in the present assumption
 to the one-loop masses.
There are solutions of the stationary conditions,
 if at least one of three vacuum expectation values is zero.

In case of $c=0$ we have
\begin{equation}
 a^{2} =
  \xi_2 - {2 \over {3g^2}} ( 2m_1^2 + m_2^2 )
  = - {1 \over 2}\phi_R + {\sqrt{3} \over 2}\phi_I
    - {2 \over {g^2}} m^2,
\end{equation}
\begin{equation}
 b^{2} =
  - \xi_1 - {2 \over {3g^2}} ( m_1^2 + 2m_2^2 )
  = {1 \over 2}\phi_R + {\sqrt{3} \over 2}\phi_I
    - {2 \over {g^2}} m^2
\end{equation}
 as a solution of the stability conditions,
 which is possible in case of
\begin{equation}
 \phi_I > 4m^2/\sqrt{3}g^2
\quad {\rm and} \quad
 |\phi_R| < \sqrt{3} \phi_I - 4m^2/g^2.
\end{equation}
There are two ``Higgs doublets'' in this case.
In case of $b=0$ we have
\begin{equation}
 a^{2} =
  - {1 \over 2} \xi_3 - {2 \over {3g^2}} ( 2m_1^2 + m_3^2 )
  = - \phi_R - {2 \over {g^2}} m^2,
\end{equation}
\begin{equation}
 c^{2} =
    \xi_1 - {2 \over {3g^2}} ( m_1^2 + 2m_3^2 )
  = - {1 \over 2}\phi_R - {\sqrt{3} \over 2}\phi_I
    - {2 \over {g^2}} m^2
\end{equation}
 as a solution of the stability conditions,
 which is possible in case of
\begin{equation}
 \phi_R < -2m^2/g^2
\quad {\rm and} \quad
 \phi_I < - \phi_R/\sqrt{3} - 4m^2/\sqrt{3}g^2,
\end{equation}
 and in case of $a=0$ we have
\begin{equation}
 b^{2} =
    {1 \over 2} \xi_3 - {2 \over {3g^2}} ( 2m_2^2 + m_3^2 )
  = \phi_R - {2 \over {g^2}} m^2,
\end{equation}
\begin{equation}
 c^{2} =
    - \xi_2 - {2 \over {3g^2}} ( m_2^2 + 2m_3^2 )
  = {1 \over 2}\phi_R - {\sqrt{3} \over 2}\phi_I
    - {2 \over {g^2}} m^2
\end{equation}
 as solutions of the stability conditions,
 which is possible in case of
\begin{equation}
 \phi_R > 2m^2/g^2
\quad {\rm and} \quad
 \phi_I < \phi_R/\sqrt{3} - 4m^2/\sqrt{3}g^2.
\end{equation}
 In these two cases there is one ``Higgs doublet''.
Each of three solutions is possible in case of that
 the values of parameters, $\phi_R$ and $\phi_I$,
 take appropriate values for positive squared vacuum expectation values.
The possible regions in $\phi_R$-$\phi_I$ space for three solutions
 are exclusive, namely,
 there is no region in which two of three or all of three are possible.
It can be shown that
 the point $a=b=c=0$ is always unstable,
 if one of three solutions are realized.
We can conclude that
 the flat direction is lifted up by mass terms,
 and there is no flat direction any more.
Three D3-branes
 are separated from the origin of the ($X^8$, $X^9$)-plane
 in some ``modified $\mathbb{Z}_{3}$'' symmetric way,
 and their positions are stabilized and fixed.

Now we can discuss the relationship
 between the scale of gauge symmetry breaking and string scale.
Define the scale of SU$(2)$ gauge symmetry breaking $v$ as
\begin{equation}
 \left(\frac{v}{\sqrt{2}}\right)^{2}
  \equiv a^2 + b^2
  = 2 \left( {\sqrt{3} \over 2} \phi_I - {2 \over {g^2}} m^2 \right)
\end{equation}
 for the two ``Higgs doublet'' case, and
\begin{equation}
 \left(\frac{v}{\sqrt{2}}\right)^{2}
  \equiv a^2 = - \phi_R - {2 \over {g^2}} m^2
\quad {\rm or} \quad
 \left(\frac{v}{\sqrt{2}}\right)^{2}
  \equiv b^2 = \phi_R - {2 \over {g^2}} m^2,
\end{equation}
 for the one ``Higgs doublet'' cases.
It is common in all these three cases that
 negative contributions by the one-loop mass
 are made to be positive by positive contributions of FI coefficients.
Since we are assuming that
 the magnitudes of $\phi_R$ and $\phi_I$ are less than string scale $M_s$,
 it is natural and conservative to consider that both contributions
 are the same orders of magnitudes.
Therefore, for two or one ``Higgs doublet'' cases
\begin{equation}
 v^2 \simeq {{8m^2} \over {g^2}} \simeq {{M_s^2 \over {2 \pi^2 \omega^2}}}
\quad {\rm or} \quad
 v^2 \simeq {{4m^2} \over {g^2}} \simeq {{M_s^2 \over {4 \pi^2 \omega^2}}},
\end{equation}
 respectively.
It is possible to obtain estimates
 of natural string scales in low-scale string models
 by substituting electroweak symmetry breaking scale, $250$ GeV, to $v$.
We obtain
\begin{equation}
 M_s \simeq 1100 \times \omega \,\, {\rm GeV}
\quad {\rm or} \quad
 M_s \simeq 1600 \times \omega \,\, {\rm GeV}
\end{equation}
 for two or one ``Higgs doublet'' cases, respectively.
Since present experimental bounds for $M_s$
 through the possible string resonances in dijet evens are
 $M_s > 4.31$ TeV by the CMS experiment \cite{CMS:2012yf}
 and $M_s > 3.61$ TeV by the ATLAS experiment \cite{ATLAS:2012pu},
 we need at least $\omega \gtrsim 2.5$ or some other mechanisms
 of gauge symmetry breaking having $m^2 < 0$, for example.

\section{Summary}

A non-supersymmetric D-brane system
 with four D3-branes and three anti-D7-branes
 at $\mathbb{C}^{3}/\mathbb{Z}_{3}$ orbifold singularity
 has been investigated.
Especially,
 the potential of D3-brane moduli field has been calculated
 by using effective field theory and superstring world-sheet theory.
We have found that
 the potential represents flat directions
 along which the original gauge symmetry on the D3-brane stack
 U$(2)\times $U$(1)_1\times $U$(1)_2$
 is spontaneously broken to U$(1)\times $U$(1)'$.
A geometrical understanding of the gauge symmetry breaking
 by the separations of D3-branes has been given
 with the essential help of the potential
 from explicit world-sheet calculations.

The effect of blowing up the singularity,
 or the effect of Fayet-Iliopoulos terms,
 which are controlled by the vacuum expectation values of
 twisted closed string states,
 have also been investigated.
The original flat direction is deformed,
 especially the origin of the moduli space becomes always unstable,
 and the gauge symmetry breaking is necessary
 with an assumption that twisted moduli fields are stabilized
 by some unspecified mechanism.
The geometrical understanding remains applicable with Fayet-Iliopoulos terms.

Since the system has no supersymmetry,
 there must be one-loop corrections to the masses of D3-brane moduli fields.
An order estimate
 based on the explicit string one-loop calculation has been given,
 though there is a difficulty of infrared divergences
 due to uncanceled tadpoles.
The deformed flat direction is lifted up by the one-loop masses,
 and there are some stationary and stable points in D3-brane moduli space
 at which spontaneous gauge symmetry breaking occurs.
The relations between
 the scale of gauge symmetry breaking and string scale have been obtained.
Naturally, the scale of gauge symmetry breaking
 is one order less than the string scale.
This result gives an indication of low-scale (or TeV scale) string models
 with experimental lower bounds on string scale.

\section*{Acknowledgements}
NK is supported in part by
 Grant-in-Aid for Scientific Research on Innovative Areas (\# 24104505)
 from the Ministry of Education, Culture, Sports, Science and Technology
 of Japan.

\end{document}